\documentclass[preprint,12pt]{elsarticle}
\usepackage{amssymb}
\journal{}
\usepackage{epsfig}
\usepackage{ifthen}
\usepackage{amsfonts}
\usepackage{amssymb}
\usepackage{amsmath}
\usepackage{amscd}
\usepackage{amsthm}
\usepackage{fancyvrb}
\usepackage{epsfig}
\usepackage{mathtools}
\usepackage{epsfig}
\usepackage{bbm}
\usepackage{fullpage}
\usepackage[boxed,algosection,linesnumbered]{algorithm2e}
\begin{document}
\begin{frontmatter}
\title{Transforming NP to P: An Approach to Solve NP Complete Problems}
\author[]{Wenhong Tian$^{a,b,c}$}
\author[1]{GuoZhong Li}
\author[1]{Xinyang Wang}
\author[1]{Qin Xiong}
\author[1]{YaQiu Jiang}
\address[1]{School of Information and Software Engineering, \\University of Electronic Science and Technology of China (UESTC)}
\address[2]{BigData Research Center of UESTC}
\address[3]{CLOUDS Lab, Dept. of Computing and Information Systems, The University of Melbourne, Australia}
\begin{abstract}
NP complete problem is one of the most challenging issues. The question of whether all problems in NP are also in P is generally considered one of the most important open questions in mathematics and theoretical computer science as it has far-reaching consequences to other problems in mathematics, computer science, biology, philosophy and cryptography. There are intensive research on proving `NP not equal to P' and `NP equals to P'. However, none of the `proved' results is commonly accepted by the research community up to now. In this paper, instead of proving either one, we aim to provide new perspective: transforming two typical NP complete problems to exactly solvable P problems in polynomial time. This approach helps to solve originally NP complete problems with practical applications. It may shine light on solving other NP complete problems in similar way. \\
\end{abstract}
\end{frontmatter}
\section{Introduction}
P versus NP problem is one of seven Millennium Prize Problems in mathematics that were stated by the Clay Mathematics Institute in 2000. As of March 2015, six of the problems remain unsolved. A correct solution to any of the problems results in a US 1,000,000 prize (sometimes called a Millennium Prize) being awarded by the institute [10]. Simply speaking, P problems mean that the class of problems can be solved exactly in polynomial time while NP (non-deterministic polynomial) problem stands for a class of problems which can not be solved in polynomial time. Intuitively, NP problem is the set of all decision problems for which the instances where the answer is ``yes" have efficiently verifiable proofs of the fact that the answer is indeed ``yes". More precisely, these proofs have to be verifiable in polynomial time by a deterministic Turing machine. In an equivalent formal definition, NP problems is the set of decision problems where the ``yes"-instances can be accepted in polynomial time by a non-deterministic Turing machine [11]. NP problems has far-reaching consequences to other problems in mathematics, biology, philosophy and cryptography.
The official statement of NP problem was given by Stephen Cook [2]. We say that a problem $A$ in NP is NP-complete when, for every other problem $B$ in NP, $B$ is easier than $A$, i.e., $B<A$. In computational complexity theory, Karp's 21 NP-complete problems are a set of computational problems which are NP-complete. In his 1972 paper [5], Richard Karp used Stephen Cook's 1971 theorem that the boolean satisfiability problem is NP-complete (also called the Cook-Levin theorem) to show that there is a polynomial time many-one reduction from the boolean satisfiability problem to each of 21 combinatorial and graph theoretical computational problems, thereby showing that they are all NP-complete. This was one of the first demonstrations that many natural computational problems occurring throughout computer science are computationally intractable, and it drove interest in the study of NP-completeness and the P versus NP problem. The complexity class P is contained in NP, and NP contains many important problems. The hardest of which are called NP-complete problems, whose solutions are sufficient to deal with any other NP problems in polynomial time. The most important open question in complexity theory, is the P versus NP problem which asks whether polynomial time algorithms actually exist for NP-complete problems and all NP problems. It is widely believed that this is not the case [3]. In 2012, 10 years later, the same poll was repeated [4]. The number of researchers who answered was 126 (83\%) believed the answer to be no, 12 (9\%) believed the answer is yes, 5 (3\%) believed the question may be independent of the currently accepted axioms and therefore is impossible to prove or disprove, 8 (5\%) said either don't know or don't care or don't want the answer to be yes nor the problem to be resolved. 
Most computer scientists seem to suspect P does not equal NP. MIT computer scientist Scott Aaronson gives informal arguments against P=NP in his post, including his philosophical argument [11]:
``If P=NP, then the world would be a profoundly different place than we usually assume it to be. There would be no special value in `creative leaps', no fundamental gap between solving a problem and recognizing the solution once it is found. Everyone who could appreciate a symphony would be Mozart; everyone who could follow a step-by-step argument would be Gauss; everyone who could recognize a good investment strategy would be Warren Buffett. It is possible to put the point in Darwinian terms: if this is the sort of universe we inhabited, why wouldn't we already have evolved to take advantage of it?" 
The important thing is that \textit{Karp showed that if any of them have efficient algorithms, then they all do}. Many of these problems arise from real-world optimization problems including Load Balancing, Minimizing Total Busy Time of Multiple Machines, Bin Packing, Hamiltonian Cycle, Chromatic Number. Researchers later extend Karp's techniques to show hundreds if not thousands of natural problems are NP-complete.\\
There are intensive research on proving `NP not equal to P' and `NP equals to P'. However, none of the `proved' results is commonly accepted by the research community yet up to now. In this paper, instead of proving either one, we aim to provide new perspective: transforming two typical NP complete problems to exactly solvable P problems in polynomial time.
In the following sections, we introduce two NP problems and show how to transform them into P problems in polynomial time.
The remaining sections are organized as follows. Minimizing makespan (MinMS) in load balance of multiple machines is discussed in Section 2. Minimizing Total Power-on Time (MinTPT) of multiple machines is proposed in Section 3. Finally we conclude in 4.
\section{Minimizing the Makespan for Load Balancing of Multiple Machines}
The well-known load balance problem [2], its formulation can be described as follows. Given a set of $m$ identical machines $M_1, M_2, \ldots, M_m$ and a set of $n$ requests (jobs), each request has a processing time $p_i$ (consider only CPU processing for example). Each machine can only host one job at any time. The sum of the process time allocated on a machine is called its workload, or simply load. Traditionally, the makespan is used to measure the workload of multiple machine. The makespan is the total length of the schedule, i.e., the time length of the start-time of the first scheduled machine and the end-time of the last scheduled machine where each job occupies the whole capacity of a machine during processing. The makespan is used to measure the load balance, which is simply the maximum load on any machine. The smaller the makespan is, the better the load balance. 
The objective of load balance is to assign each request to one of machines so that the loads placed on all machines are balanced (or the maximum load on all machines is minimized). This is called offline load balance since the scheduler knows all requests and status of all machines. \textit {In this paper, requests and jobs are used interchangeably.} In Tian et al. [8], an approximation algorithm called Prepartition with near optimal results is proposed. In this paper, we provide exactly solution in polynomial time. 
Firstly we provide a formal defition of the makespan as follows. \\
\noindent \textbf{The makespan}: \textit{In any allocation of requests to machines, we can let $A(i)$ denote the set of requests allocated to machine $M_i$, under this allocation, machine $M_i$ will have total load,
\begin{equation}
L_i=\sum_{j\in A(i)}p_j
\end{equation}
where $p_j$ is the length of processing time of request $j$. And
\begin{equation}
makespan=\max_{1\leq i \leq m} L_i
\end{equation}
}
The goal of load balancing is to minimize the maximum load (makespan) on any machine.
\noindent \textbf{ Theorem 1}. \textit{Minimizing the makespan (MinMS) of offline scheduling in general case is NP-complete.}
\\
Proof: We sketch a brief proof as follows. 
We show that a well-know NP complete problem, Subset Sum problem (SSP) is polynomial time reducible to MinMS problem. Thus consider an instance of SSP with numbers $w_1, w_2, \ldots, w_n$, which corresponding to the CPU load of $n$ requests and have total CPU load W. To be load-balanced, in an ideal case, it is to let all $m$ machines have same share of total CPU load, i.e., $W/m$. This needs all allocations on all machines to be satisfied. Suppose there are $j$ jobs on machine $M_i$, this requires that $L_i$=$\sum_{j\in A(i)}p_j$=$W/m$. MinMS problem is reducible to SSP problem in polynomial time and SSP is well-known NP complete problem.
On the other hand, if there is a solution to Subset Sum problem for a given set of requests, there exists a schedule for the given set of requests. Since SSP is NP-hard in the strong sense, our problem is also NP-hard. In this way, we have found that that the MinMS problem is NP-complete problem. This completes the proof. \\
Since Minimizing the makespan (MinMS) is NP-complete and traditionally the exact solution costs exponential time, so that approximation solution is widely studied. One of pioneering work is by Graham [2], where a $\frac{4}{3}$-approximation algorithm called Longest Process Time first (LPT) is proposed, and this is the best-known approximation up to now.
In the following, we introduce a new approach to solve MinMS problem exactly in polynomial time.
Let us consider there are $m$ machines and denote OPT as the ideal (theoretical) optimal solution for a given set of $n$ requests. Firstly define
\begin{equation}
OPT=\frac{\sum_{j} p_j}{m}
\end{equation}
The ideal load balance is to allocate each machine with load exactly of OPT. It is impossible to achieve this without additional techniques. In the following, we use Virtual Machines (VMs) and apply Live VM Migration technique to achieve ideal load balance.\\
\textbf{Virtual Machines (VMs).} \textit{In computing, a virtual machine (VM) is an emulation of a particular computer system. Virtual machines operate based on the computer architecture and functions of a real or hypothetical computer, and their implementations may involve specialized hardware, software, or a combination of both.} \\
\textbf{Live VM Migrations.} \textit{Live migration refers to the process of moving a running virtual machine or application between different physical machines without disconnecting the client or application. Memory, storage, and network connectivity of the virtual machine are transferred from the original guest machine to the destination.}\\
The virtual machines and VM migrations are currently widely adopted for workload consolidation and load balance in cloud computing, BigData, etc.. Our proposed algorithm will apply live VM migrations.
Algorithm.~2.1 shows the pesudocodes of our proposed PAM algorithm. The algorithm firstly computes optimal balance value (OPT) by equation (3). It sorts all jobs in non-increasing order of their process time, and then finds a machine with the lowest makespan (load) and available capacity to allocate in sorted order, and updates the load on each machine. After all requests are allocated, the algorithm computes the makespan of each machine and finds machines with load more than OPT and machines with load less than OPT. PAM sorts additional load (the load more than OPT) in non-increasing order in those machines with load more than OPT, and sorts machines with load less than OPT in non-decreasing order. Then it partition the load more than OPT in those machines and allocate to compensate load in machines with load less than OPT to make their load exactly equal to OPT, untile all machines have load exactly equal to OPT. For practice, the scheduler has to record all partitions and their hosting machines of each request so that migrations of machines can be conducted in advance to reduce overheads.
\begin{algorithm}[!tb]
\SetArgSty{textnormal}
\caption{Partition and Migration (PAM) Algorithm}\label{PrepartitionOn }
\KwIn{VM requests come one by one indicated by their information (required VM type IDs, start times, ending times, requested capacity), $MS_i$ is the makespan of request $i$}
\KwOut{Assign a machine ID to all requests and their partitions}
Initialization: computes OPT value and allocates jobs to machines with the lowest load in longest process time (LPT) first fashion;
\ForAll{ machines with load more than OPT} {\nllabel{OPT Outer Loop Begin Line}
Sorts them in non-increasing order of load on each machine\;
Starts from the machine with largest load and the machine with least load\;
Partitions the load more than OPT on these machines to compensate load on machines with load less than OPT to make their load exactly equal to OPT.
Until all machines have load exactly equal to OPT\;
Update the load of the machine\;
}\nllabel{PrepartitonOn Outer Loop End Line}
Compute the load of each machines and total partitions (migrations)\;
\end{algorithm}
\noindent \textbf {Theorem 2}: \textit{The computational complexity of PAM is $O(nlogm)$ using priority queue data structure where $n$ is the number of requests after paritions and $m$ is total number of machines used}. \\
\noindent Proof: The priority queue is designed such that each element (machine) has a priority value (i.e.,makespan), and each time the algorithm needs to select an element from it, the algorithm takes the one with the highest priority (the smaller makespan value is, the higher priority it is). Sorting a set of $n$ number in a priority queue takes $O(n)$ time and a priority queue performs insertion and the extraction of minima in $O(logm)$ steps for $m$ machines (detailed proof of the priority queue is shown in \cite{Kleinberg2005}). Also PAM sorts the load in $m$ machines in non-increasing order and this takes $O(mlogm)$ time. Therefore, by using priority queue or related data structure, the algorithm can find a machine with the lowest makespan in $O(logm)$ time. Altogether, for $n$ requests and $m$ machines, PAM has time complexity $O(nlogm)$ where normally $n>m$. \\ 
\noindent \textbf { Theorem 3}: \textit{PAM is exact and optimal}.\\
\noindent Proof: We sketch the proof as follows.
Because OPT can be computed in advance, applying PAM will migrate the load on machines with load more than OPT to machines with load less than OPT, and make them have load exactly equal to OPT, this will bring ideal load balance to each machine. This finishes the proof. \\
\\
\noindent \textbf {Example-1}: Consider the worst case for LPT given in [2]. There are $m$ machines, and $n$=2$m$+1 jobs with process time of $m$,$m$,$m$,$m$+1,$m$+1,$m$+2,$m$+2,...,$m$+($m$-1),$m$+($m$-1). In this case, the total load of all jobs is 3$m^2$, and OPT is 3$m$. The LPT algorithm has result 4$m$-1, so its approximation ratio is $\frac{4m-1}{3m}$$\approx$$\frac{4}{3}$, as $m$ increases to very large. For PAM, OPT (3$m^2$) is computed firstly, and applying LPT, the max load of all machines is 4$m$-1 and other machines will have load 3$m$-1 . PAM partitions the load in the machine with load more than OPT (in this case $m$-1) into ($m$-1) equal parts and migrate to each of other $m$-1 machines with load less than OPT. Therefore each machine will have load exactly equal to 3$m$, which is OPT. \\
The virtual machine migration is widely adopted for workload consolidation and load balance in cloud computing [8]. By applying PAM, the original NP-complete MinMS problem is transferred to P problem and is solved in polynomial time.
\section{Minimizing Total Power-on Time of Multiple Machines with Fixed Process Intervals}
Minimizing total power-on time (MinTPT) of all machines is based on the following assumptions and definitions.\newline
1). All data are deterministic unless otherwise specified, the time is considered in discrete slotted window format. We partition the total time period [0, T] into slots of equal length ($l_0$) in discrete time, thus the total number of slots is $S$=$T/l_0$ (always making it a positive integer). The start-time of the system is set as $s_0$=0. Then the interval of a request $i$ can be represented in slot format as a tuple with the following parameters: [StartTime, EndTime, RequestedCapacity]=$[s_i, e_i, d_i]$. With both start-time $s_i$ and end-time $e_i$ are non-negative integers. We set $d_i$=1 (called unit demand) in this paper. \\
2). For all jobs, there are no precedence constraints other than those implied by the start-time and end-time. Preemption is not considered.\\
MinTPT belongs to fixed interval processing problems, see [7] for a detailed discussion about this problem. In Tian et al. [9], an approximation algorithm called MFFDE with approximation ration 3 is proposed. In this paper, we aim to solve it exactly in polynomial time.\\
\textbf{Definition 1}. \textit{The Interval Length: given a time interval $I_i$= [$s_i$, $t_i]$ where $s_i$ and $t_i$ are the start slot and end slot, the length of $I_i$ is $|I_i|$=$t_i$-$s_i$. The length of a set of intervals $I$=$\bigcup_{i=1}^{k}I_i$, is defined as $len(I)$=$|I|$=$\sum_{i=1}^{k}|I_i|$, i.e., the length of a set of interval is the sum of the length of each individual interval.}
\\
\textbf{Definition 2}. \textit{The Interval Span: The span of a set of intervals, $span(I)$, is defined as the length of the union of all intervals considered.}
\\
\textbf{Example-2}: if $I$=$\{[1,4],[2,4],[5,6]\}$, then $span(I)$=$\mid$[1,4]$\mid$+$\mid$[5,6]$\mid$=(4-1)+(6-5)=4, and $len(I)$ =$\mid$[1,4]$\mid$+$\mid$[2,4]$\mid$+$\mid$[5,6]$\mid$=6. Note that $span(I)\leq len(I)$ and equality holds if and only if $I$ is a set of non-overlapping intervals.\\
The MinTPT problem has the following formulation: the input is a set of $n$ jobs (requests) $J$= ${j_1,...,j_n}$. Each job $j_i$ is associated with an interval $[s_i,e_i]$ in which it should be processed, where $s_i$ is the start-time and $e_i$ the end-time, both in discrete time. Set $p_i$=$e_i$-$s_i$ as the process time of job $j_i$. For the sake of simplicity, we concentrate on CPU-intensive applications and consider CPU capacity only. The capacity parameter $g \geq 1$ is the maximal CPU capacity a single machine provides. Each job requests unit capacity $d_i$=1. The power-on time of $M_i$ is denoted by its working time interval length $b_i$. The optimizing objective is to assign the jobs to machines such that the total power-on time of all machines is minimized. Note that the number ($m\geq 1$) of machines to be used is part of the output of the algorithm and takes integer value. This problem is called MinTPT problem for abbreviation.\\
\textbf{THEOREM 4}. \textit{MinTPT problem is NP-complete problem in general case.}\\
\textbf{Proof}: 
For completeness, we sketch the proof as follows. $K$-PARTITION problem is NP-complete [6]. For a given arrangement $P$ of positive numbers and an integer $k$, $K$-PARTITION problem is to partition $P$ into $K$ ranges so as the sums of all the ranges are close to each other. $K$-PARTITION problem can be reduced to our MinTPT problem as follows. For a set of jobs $J$, each has capacity demand $d_i$=1 , partitioning $J$ by their capacities into $K$ ranges, is the same to allocate $K$ ranges of jobs with capacity constraint $g$ (i.e. the sum of each range is at most $g$). On the other hand, if there is a solution to $K$-PARTITION for a given set of intervals, there exists a schedule for the given set of intervals. Since $K$-PARTITION is NP-hard in the strong sense, our problem is also NP-hard. In this way, we have found that that the MinTPT problem is NP-complete problem.\\
\\
\textbf{THEOREM 5}: \textit{The lower bound (OPT) for MinTPT problem is the sum of the minimum number of machines used in each slot, i.e., the lower bound is to allocate exactly min number of machines needed to each time slot}. \\
\textbf{Proof:} MinTPT problem is offline scheduling, for a given set of jobs $J$, we can find the minimum number of machines needed for each time slot, denoted as $l_1, l_2,...l_k$ for total $k$ time slots under consideration, where $l_i$ is the minimum number of machines needed for time slot $i$. By the definition of the interval span and power-on time of each machine, $OPT(I)=\sum_{i=1}^{S} \lceil \frac{L_i}{g} \rceil$=$\sum_{i=1}^{S} l_i$, here $L_i$ is the sum of load for time slot $i$. The total power-on time of all machines is the sum of min number of machines in all time slots in this way, i.e., the lower bound is the sum of the minimum number of machines used in each slot. This is the minimum total power-on time of all machines. This completes the proof. \newline
\textbf{Remark$\#1$}: The theoretical lower bound given in THEOREM 1 is not easy to achieve if each request has to be processed on a single machine without migration. Finding a subset of jobs for each machine to minimize total power-on time is known to be NP-complete [10]. 
\\
\begin{figure} [htp!]
\begin{center}
\hfill
{\includegraphics [width=0.53\textwidth, angle=0] {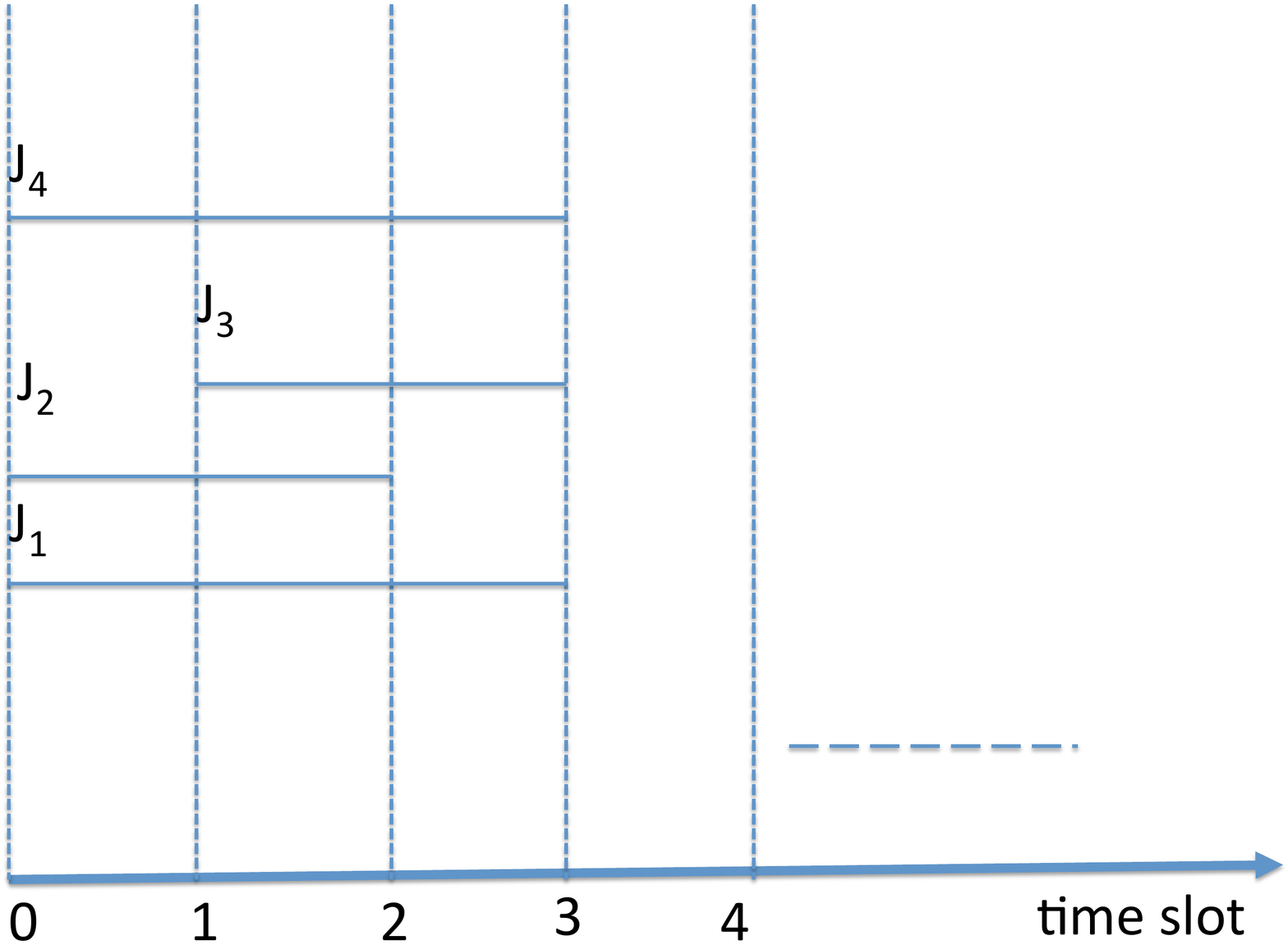}}
\hspace*{\fill}
\caption{A simple example of MinTPT}
\end{center}
\end{figure}
\textbf{Example-3:} As shown in Fig.~1, consider there are 4 job requests and $g$=3, Jobs $J_1$, $J_2$, $J_3$, $J_4$ have start-time, end-time and capacity demand [0, 3, 1], [0, 2, 1], [1, 3, 1], [0, 3, 1] respectively. The min number of machines needed is 1, 2, 1 respectively in three time slots and total power-on time is 4 in theoretical lower bound (OPT). Without job migration, one solution is to allocate $J_1$, $J_2$ and $J_4$ to one machine by earliest start time first and $J_3$ to another machine; in this way the actual total number of machines needed is 2 and the total power-on time is 5. With job migration, one can allocate $J_1$, $J_2$ and $J_4$ to one machine ($m_1)$ during interval [0, 3], allocate $J_3$ to another machine ($m_2$) during interval [1, 2] and migrate $J_3$ to $m_1$ during interval [2,3]; in this way, the total power-on time is 4, equal to the OPT. \newline
\textbf{THEOREM 6}. \textit{MinTPT problem obtains optimum result if job migration is allowed.}\\
\textbf{Proof}: From THEOREM 5, we know that there is a theoretical lower bound for MinTPT problem. The MinTPT as proved in THEOREM 4, is NP-complete in general case without job migration. However with job migration, a job can be migrated from one machine to another machine to be continuously proceeded, it is possible to obtain the lower bound. The method is introduced in Algorithm 3.~1 OPT-With-Migration. Algorithm 3.1 firstly sorts all jobs in non-decreasing order of jobs' start-time (line 1) and represents load of each slot by the min number of machines needed (line 3-4); it allocate jobs non-decreasing order of their start-time (line 6-7); and migrates the job to an existing machine when the min number of machines will be more than slot load (line 10-11); it updates load of each machine (line 12). In this way, the algorithm obtains the theoretical lower bound (denote as OPT in this paper) with certain number of migrations. This completes the proof. \\
\\
\begin{algorithm}[tb]
\SetArgSty{textnormal}
\caption{Algorithm 1 Lower Bound Migration (LBM)}\label{ OPT-With-Migration }
\KwIn{A Job instance $J$=$\{j_1, j_2, \ldots, j_n\}$, and $g$, the max capacity $g$ of a machine}
\KwOut{The scheduled jobs and total power-on time of all machines}
Sort all jobs in non-decreasing order of their start-time ($s_i$ for job $i$), such that $s_1\leq s_2...\leq s_n$, set $h$=1\;
\ForAll{slots under consideration} {\nllabel{OPT Outer Loop Begin Line}
Consider unit demand, represent load of slot $i$ by the min number of machines needed for it, denoted as $l_i$ (take integer value by ceiling function)} \
\ForAll{jobs under consideration} {\nllabel{OPT Outer Loop Begin Line}
Sorting in non-decreasing order of jobs' start-time, i.e., earliest start-time first (ESTF), as a trial, allocate to the first machine, open a new machine and set $h$=$h$+1 if needed\;
}\nllabel{OPT Outer Loop End Line}
\ForAll{ slot $i$ from 1 to $S$ }{ \nllabel{LCS First Inner Loop Begin Line}
\If{ $l_i$ is not reached in the slot}{
\textbf {else} {the min number of machines is more than $l_i$,}{\nllabel{second for}
one or more later allocated jobs are migrated to an existing machine which still can host in the slot, so that the min number of machines in slot $i$ is equal to $l_i$.
} \nllabel{end If}
} \nllabel{OPT First Inner Loop End Line}
fix the allocation and update load of $M_h$\;
} \nllabel{OPT Second Inner Loop End Line}
Find power-on times of all machines\;
Return the set of machines used, and the total power-on time of all machines\
\end{algorithm}
\textbf{THEOREM 7}. \textit{LBM has computational complexity of $O(nlogn)$+$O(S)$ where $n$ is the number of jobs and $S$ is the toal number of slots.}\\
\textbf{Proof}: LBM firstly sorts all jobs in non-decreasing order of their start-time, this takes $O(nlogn)$ time, and it considers $S$ slots for possible migration, taking $O(S)$ time. So all together LBM has computational complexity of $O(nlogn)$+$O(S)$.\\
By applying limited number of migrations, the original NP complete problem MinTPT can be transformed into P problem. 
\section{Conclusions and Future Work}
As we discussed in previous sections, transforming the MinMS and MinTPT problems into exactly solvable ones in polynomial time have some common features. Their OPT value can be computed in advance so that partition and migration can be applied to achieve the OPT results.
How about other NP complete problems? Can they also be solved in  similar way? In our previous work [12], we ever proved that the NP complete problem, traveling salesman problem (TSP) can be solved with near optimal results by iterative k-opt algorithm. However, 
whether TSP can be transformed into P problem is not known yet. According to Karp's result [5] that if any of NP problems have efficient algorithms, then they all do. Our proposed approach can shine light on other NP problems. In the near future, we will look at  other NP complete problems and to study the possibility of transforming them into P problems. 
\section*{Acknowledgments}
This research is sponsored by the National Natural Science Foundation of China (NSFC) (Grand Number:61450110440). Dr. Wenhong Tian is currently a visiting fellow at CLOUDS lab at the University of Melbourne, Australia.
\bibliographystyle{elsarticle-num}

\end{document}